\documentclass[12pt]{article}

\usepackage{authblk}
\usepackage[utf8]{inputenc}
\usepackage{cite}
\usepackage{graphicx}
\usepackage{color,soul}

\title{Magnetotransport properties of La$_{1-x}$Ca$_x$MnO$_3$ (0.52 $\leq x \leq$ 0.75): signature of phase coexistence}
\author[a]{M. Čulo%
\thanks{Electronic address: \texttt{mculo@ifs.hr}; Corresponding author}}
\author[b]{M. Basletić}
\author[b]{E. Tafra}
\author[b]{A. Hamzić}
\author[a]{S. Tomić}
\author[c]{F. Fischgrabe}
\author[c]{V. Moshnyaga}
\author[a]{B. Korin-Hamzić}
\date{}
\affil[a]{Institut za fiziku, P. O. Box 304, HR-10001 Zagreb, Croatia}
\affil[b]{University of Zagreb, Faculty of Science, Department of Physics, P. O. Box 331, HR-10001 Zagreb, Croatia}
\affil[c]{I. Physikalisches Institut, Georg-August-Universit\"{a}t G\"{o}ttingen, Friedrich-Hund-Platz 1, 37077 G\"{o}ttingen, Germany}

\begin{document}
\maketitle

\newpage

\begin{abstract}
We report the temperature and magnetic field dependence of transport properties in epitaxial films of the manganite La$_{1-x}$Ca$_{x}$MnO$_{3}$ in the overdoped region of the phase diagram for $x > 0.5$, where a charge--ordered (CO) and an antiferromagnetic (AF) phase are present. Resistivity, magnetoresistance and angular dependence of magnetoresistance were measured in the temperature interval $4.2 ~\mathrm{K} < T < 300 ~\mathrm{K}$, for three concentrations $x = 0.52, 0.58$ and $0.75$ and in magnetic fields up to 5 T. The semiconductor/insulator--like behavior in zero field was observed in the entire temperature range for all three concentrations \textit{x} and the electric conduction, at lower temperatures, in the CO state obeys 3D Mott's variable--range hopping model. A huge negative magnetoresistance for $x = 0.52$ and $x = 0.58$, a metal--insulator transition for $B > 3 ~\mathrm{T}$ for $x = 0.52$ and the presence of anisotropy in magnetoresistance for $x = 0.52$ and $x = 0.58$ show the fingerprints of colossal magnetoresistance (CMR) behavior implying the existence of ferromagnetic (FM) clusters. The declining influence of the FM clusters in the CO/AF part of the phase diagram with increasing $x$ contributes to a possible explanation that a phase coexistence is the origin of the CMR phenomenon.
\end{abstract}

Keywords: manganites,  charge order, magnetotransport, variable--range hopping, CMR, phase coexistence

\section{Introduction}

The mixed–valence perovskites La$_{1–x}$Ca$_x$MnO$_3$ belong to the family of manganese oxides, widely known as manganites, with a general chemical formula $R_{1–x}A_{x}$MnO$_{3}$, where $R$ usually stands for a trivalent rare earth and $A$ for a divalent alkaline earth ion. A major interest for manganites started when a huge decrease of resistivity in external magnetic field, i.e. a magnetically induced metal–insulator phenomenon, known as colossal magnetoresistance (CMR), was found offering an important possibility of application (for general reviews see Refs. \cite{ramakrishnan,dagotto05,dagotto01,tokura,imada} and references therein).

The unique properties of mixed valence manganites are determined by complex spin, charge, and orbital ordered phases and their coexistence and/or competition results in a rich and complex phase diagram. The underlying physics is based on a complexity of electronic interactions of comparable energy scales superimposed on electron–lattice interactions that give rise to a local intrinsic inhomogeneity in the electronic phases. Such intricate coupling leads to interesting properties such as CMR, charge ordering (CO), multiferroicity, and electronic phase separation. A particularly important material parameter, that causes transitions among distinct phases, is the chemical doping of the parent manganite. Upon doping by a divalent element, the Mn is driven into a mixed valence state Mn$^{3+}$/Mn$^{4+}$ and the physical properties vary with the relative concentration of Mn$^{3+}$ and Mn$^{4+}$. The original cation and the dopant have, in general, different sizes, and they are distributed randomly in the structure and as a consequence, such systems are characterized by local distortions.

The La$_{1–x}$Ca$_{x}$MnO$_{3}$ compound is a well known system showing a metal–\newline insulator transition and CO phenomenon at different doping levels. For $0.2 \leq x \leq 0.5$ the compound is ferromagnetic (FM) and metal–like below the Curie temperature $T_{p}$, and paramagnetic and insulator--like above $T_{p}$. A huge negative magnetoresistance (CMR) occurs near $T_{p}$. For $0.5 \leq x \leq 0.87$ the compound is paramagnetic and insulator--like above the charge ordering transition ($T_{\mathrm{CO}}$) and becomes a charge–ordered insulator at lower temperatures. Above $T_{\mathrm{CO}}$, Mn$^{3+}$ and Mn$^{4+}$ ions are randomly distributed within the MnO$_{2}$ plane in the lattice, while the ordering of Mn$^{3+}$ and Mn$^{4+}$ ions within the MnO$_{2}$ plane below $T_{\mathrm{CO}}$ leads to an exotic static stripe phase with an insulating antiferromagnetic (AF) ground state \cite{mori,radaelli}.

At present, it is believed that one of the key features of manganites is their intrinsic electronic inhomogeneity consisting of spatial regions with different electronic orders i.e. the coexistence of competing ferromagnetic and antiferromagnetic/paramagnetic phases. This phenomenon is generally called a ``phase separation'' \cite{dagotto01,dagotto_science,nagaev}. Moreover, the possibility of the phase inhomogeneity or the phase separation has often been argued as the essential ingredient in CMR physics. However, it is also important to note that due to a technical procedure in the sample preparations the appearance of extrinsic inhomogeneities (in chemical composition, in structure, strain induced etc.) in all manganites cannot be neglected and may influence the phase separation as well \cite{dagotto_science,belevtsev,gosnet}. 

In our study we focused on the much less investigated part of the phase diagram for $x > 0.5$ considering the importance of understanding the behavior of the insulating phase. Here we have to point out that transport properties of the La$_{1-x}$Ca$_{x}$MnO$_{3}$ ($0.5 \leq x < 1$) polycrystalline samples were studied in magnetic fields up to 14 T by other authors \cite{zhou} and it was shown that for $x = 0.5$ the resistivity shows metal–like behavior for $B > 12$ T, while for higher concentrations this behavior was not found up to 14 T. Note that when these materials are prepared in the form of thin films (which typically show higher crystalline order) their properties can be different with respect to the bulk materials (that are polycrystalline samples typically showing the crystalline disorder) \cite{belevtsev,gosnet,aydogdu,schmidt}. In short, we will mention the most important results obtained on La$_{1–x}$Ca$_{x}$MnO$_{3}$ manganites related to bulk material and films. In the bulk material for $x < 0.5$ (where the comparison with films was investigated) and at the metal-insulator transition (MIT) a resistivity increase is expected because the charge carriers are scattered by grain boundaries \cite{egilmez2011}. On the other hand, the MIT temperatures in bulk materials are lower than in thin film samples (of the same composition) suggesting that the epitaxial strain in the films modifies the in-plane and out-of-plane hopping amplitudes that consequently enhance the ferromagnetic transition temperature \cite{perroni}. However, in a film, the lattice distortion can be tuned with choice of substrate or film thickness. It is interesting that for $x = 0.5$ and different substrates the significant difference in magnetotransport properties was found showing higher or lower magnetoresistance at all temperatures for certain substrates and/or film thickness \cite{aydogdu,egilmez2009,zarifi,guitierrez,zhang}. This could be explained as the result of the different strain states the films experience because of the lattice mismatch between the substrate and the material. The lattice strain decreases with an increasing thickness. The different strain states modify the orbital states and consequently the magnetotransport properties of these samples \cite{tokura2000,okimoto}. Also, they can influence the phase separation phenomena \cite{gosnet,aydogdu}. The resistivity and the MIT temperature of the films for $x < 0.5$ were found to decrease with a decreasing film thickness while a reduction in the film thickness increases the magnetoresistance of these films as well as the anisotropy of magnetoresistance \cite{egilmez2011}. At this point, it is also important to mention that the optical properties study was performed comparing bulk material and films in overdoped region for $x > 0.5$, with the additional motivation to clarify the role of disorder in the physics of overdoped La$_{1-x}$Ca$_{x}$MnO$_{3}$ manganites \cite{gorshunov}. The films in this study were prepared in the same way as the films in our study \cite{moshnyaga}. The results confirmed that  disorder effects are more pronounced in bulk material than in films as well as that disorder must be taken into account when analyzing low-energy properties.

The aim of our work is to study magnetotransport properties for a series of samples prepared as thin films, in the poorly investigated overdoped part of the phase diagram for $0.5 < x \leq 0.75$. The idea was to follow the possible changes associated with the existence and/or absence of the CMR phenomenon. For this purpose we performed magnetotransport measurements on thin films, prepared in the same way on the same substrate and with the same thickness, for three different concentrations: $x = 0.52$ close to the boundary between FM and CO/AF state, $x = 0.75$ deep in the CO/AF state and $x = 0.58$ in between.

\section{Experiment}

La$_{1–x}$Ca$_{x}$MnO$_{3}$ thin films have been prepared on MgO(100) substrates by a metalorganic aerosol deposition technique \cite{moshnyaga}. The films are 170 nm thick and have typical dimensions $5 – 10 \times 1 ~\mathrm{mm}^2$. The planes of the films correspond to the crystallographic \textit{ac} plane of the orthorhombic $P_{nma}$ structure. However, the samples show twinning since \textit{a} and \textit{c} lattice constants are almost the same. Therefore, it is not possible to associate the principal axes of the films to the crystallographic \textit{a} and \textit{c} directions. The close proximity of the \textit{c} lattice constants of films and bulk samples indicate a very small amount of stress in the films under investigation \cite{moshnyaga}.

We performed detailed measurements of resistivity as a function of temperature, magnetic field and the angle between current and the magnetic field on the La$_{1–x}$Ca$_{x}$MnO$_{3}$ thin films for three different concentrations $x = 0.52$, $x = 0.58$ and $x = 0.75$. The resistivity was measured by the standard four–contact \textit{dc} technique using currents between 1 nA and 10 $\mu$A in the temperature interval 4.2–300 K and magnetic fields up to 5 T. The current was applied along the long axis of the sample. 

The field dependent resistivity was measured by changing the magnetic field strength \textit{B} at fixed temperatures with \textit{B} perpendicular to the film plane. Magnetoresistance (MR) is defined as: $\mathrm{MR} = \Delta\rho/\rho_{0} = [\rho(B)–\rho(0)]/\rho(0)$ where $\rho(B)$ denotes resistivity in magnetic field, and $\rho(0)$ the zero–field resistivity. The angular dependent resistivity was measured at 5 T and at fixed temperatures by changing the angle $\theta$ between the current and the magnetic field. For $\theta = 0^{\circ}$ the magnetic field is perpendicular, and for $\theta = 90^{\circ}$ the magnetic field is parallel to the current, respectively. Anisotropy of magnetoresistance (AMR) is defined as: $\mathrm{AMR} = [\rho(5 ~\mathrm{T},\theta)–\rho(5 ~\mathrm{T},\theta=0^{\circ})]/\rho(5 ~\mathrm{T},\theta=0^{\circ})$, where $\rho(5 ~\mathrm{T},\theta)$ denotes the resistivity at an angle $\theta$ and $\rho(5 ~\mathrm{T}, \theta = 0^{\circ})$ the resistivity at the angle $\theta = 0^{\circ}$ (i.e. the field perpendicular to the current and the sample plane), both in the magnetic field 5 T. Defined this way, the AMR represents the anisotropy between transversal and longitudinal magnetoresistance.

\section{Results and discussion}

\begin{figure}
\centering
\includegraphics[width=0.7\textwidth]{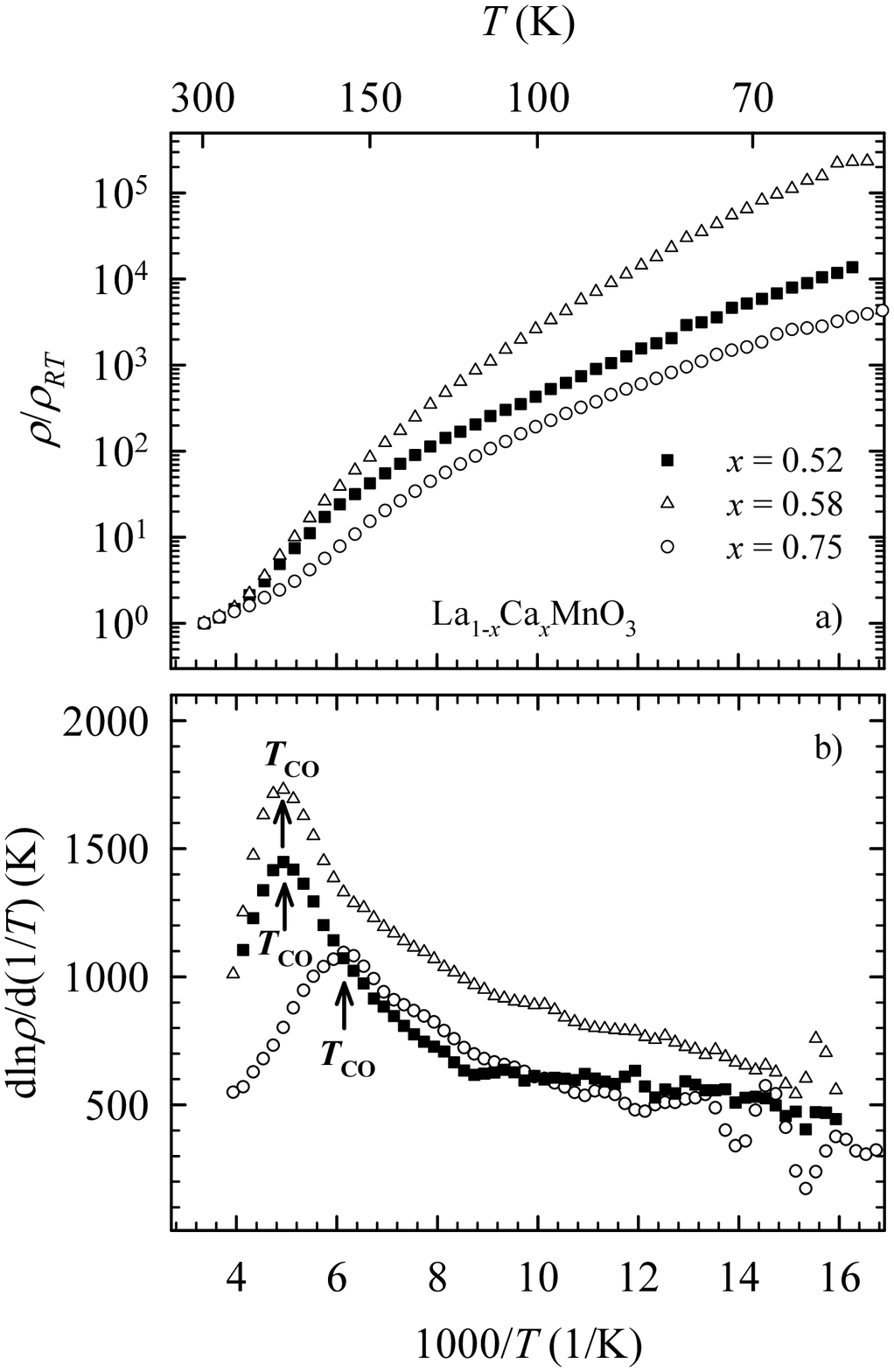}
\caption{(a) The normalized resistivity $\rho/\rho_{RT}$ versus 1/$T$ for La$_{1–x}$Ca$_{x}$MnO$_{3}$, $x = 0.52$, $x = 0.58$ and $x = 0.75$. (b) The derivative d(ln$\rho$)/d(1/$T$) vs. 1/$T$. The arrows indicate the charge order transition temperature $T_{\mathrm{CO}}$.}
\label{fig:resistivity}
\end{figure}

Figure \ref{fig:resistivity}a shows the temperature dependence of the resistivity $\rho(T)$ for La$_{1–x}$Ca$_{x}$MnO$_{3}$ in the temperature range $50 ~\mathrm{K} < T < 300 ~\mathrm{K}$, for the three different values of \textit{x} which amount 0.52, 0.58 and 0.75. The resistivity values are normalized to room temperature values $\rho_{RT}$ in order to present more clearly the difference in $\rho$ vs. $1/T$ behavior for various \textit{x}. The measured room temperature resistivity values are $\rho_{RT} = 8 ~\mathrm{m\Omega cm}$, $\rho_{RT} = 5 ~\mathrm{m\Omega cm}$ and $\rho_{RT} = 18 ~\mathrm{m\Omega cm}$ for $x = 0.52$, 0.58 and 0.75, respectively (see Table \ref{tab}). Semiconductor/insulator--like behavior is observed in the entire temperature range for all three concentrations $x$, as expected for $x > 0.5$. The signature of the charge ordering phase transition $T_{\mathrm{CO}}$ is reflected as a change in the slope of the resistivity and the $T_{\mathrm{CO}}$ values are identified as the positions of the maxima in the plots of logarithmic resistivity derivative d(ln$\rho$)/d($1/T$) vs. inverse temperature, amounting 200~K, 210 K and 160 K for $x = 0.52$, 0.58 and 0.75, respectively (see Fig. \ref{fig:resistivity}b and Table \ref{tab}). These $T_{\mathrm{CO}}$ values show satisfactory agreement with the values presented in the well known phase diagram \cite{schiffer,millis}. In the high temperature region ($T > T_{\mathrm{CO}}$) and above the room temperature a common view is that carriers form small dielectric polarons \cite{coey,worledge}. The hopping motion of polarons leads to a resistivity of the form $\rho(T)= AT\mathrm{exp}(E_{A}/k_{B}T)$, where \textit{A} is a constant and $E_{A}$ is the activation energy, i.e., the potential barrier that the polaron must surmount in order to hop into the next site. The data above $T_{\mathrm{CO}}$ and up to the room temperature cover a narrow temperature interval and the analysis should be taken with some caution. We have nevertheless estimated the  \textit{A} and $E_{A}$ values (see Table \ref{tab}), and they agree quite well with the data of other authors \cite{worledge}. First, the conductivity prefactor $1/A$ decreases with increasing concentration \textit{x} (for $x > 0.5$) following the theoretical prediction for nearest–neighbor hopping when only on–site Coulomb repulsion (and no other correlations) is taken into account. Also, the activation energy $E_{A}$ decreases monotonically as \textit{x} is increased, which is in accordance with a suggestion that the more highly charged La$^{3+}$ ions bind polarons more tightly than Ca$^{2+}$ ions. 

The analysis of our resistivity data below $T_{\mathrm{CO}}$ shows that log$\rho$ plotted against $1/T$ is not a straight line, i.e. that the electric conduction is not a simple thermally activated process like in conventional semiconductors. Note that the random distribution of La and Ca ions in the system may cause a variable--range hopping mechanism that usually appears in semiconductors where dopant atoms are randomly distributed \cite{mott}. 
\begin{table}
\centering
\begin{tabular}{|l|c|c|c|} \hline
{}&$x = 0.52$&$x = 0.58$&$x = 0.75$\\ \hline
$\rho_{RT}$(m$\Omega$cm)&$8 \pm 1$&$5 \pm 1$&$18 \pm 2$\\ \hline
$T_{\mathrm{CO}}$(K)&$200 \pm 10$&$210 \pm 10$&$160 \pm 10$\\ \hline
$E_{A}$(meV)&$99 \pm 6$&$82 \pm 9$&$69 \pm 3$\\ \hline
$1/A$(K/$\Omega$cm)&$(2.0 \pm 0.5) \cdot 10^{6}$&$(1.5 \pm 0.4) \cdot 10^{6}$&$(2.4 \pm 0.3) \cdot 10^{5}$\\ \hline
$T_{0}$($10^{7}$ K)&$4.7 \pm 0.7$&$13 \pm 3$&$3.5 \pm 0.5$\\ \hline
$\rho_{\infty}$($\Omega$cm)&$(1.5 \pm 0.4) \cdot 10^{-11}$&$(3.0 \pm 0.5) \cdot 10^{-14}$&$(1.0 \pm 0.3) \cdot 10^{-10}$\\
\hline
\end{tabular}
\caption{\textit{dc} transport parameters of La$_{1–x}$Ca$_{x}$MnO$_{3}$ for $x = 0.52$, $x = 0.58$ and $x = 0.75$.}
\label{tab}
\end{table}

\begin{figure}
\centering
\includegraphics[width=0.7\textwidth]{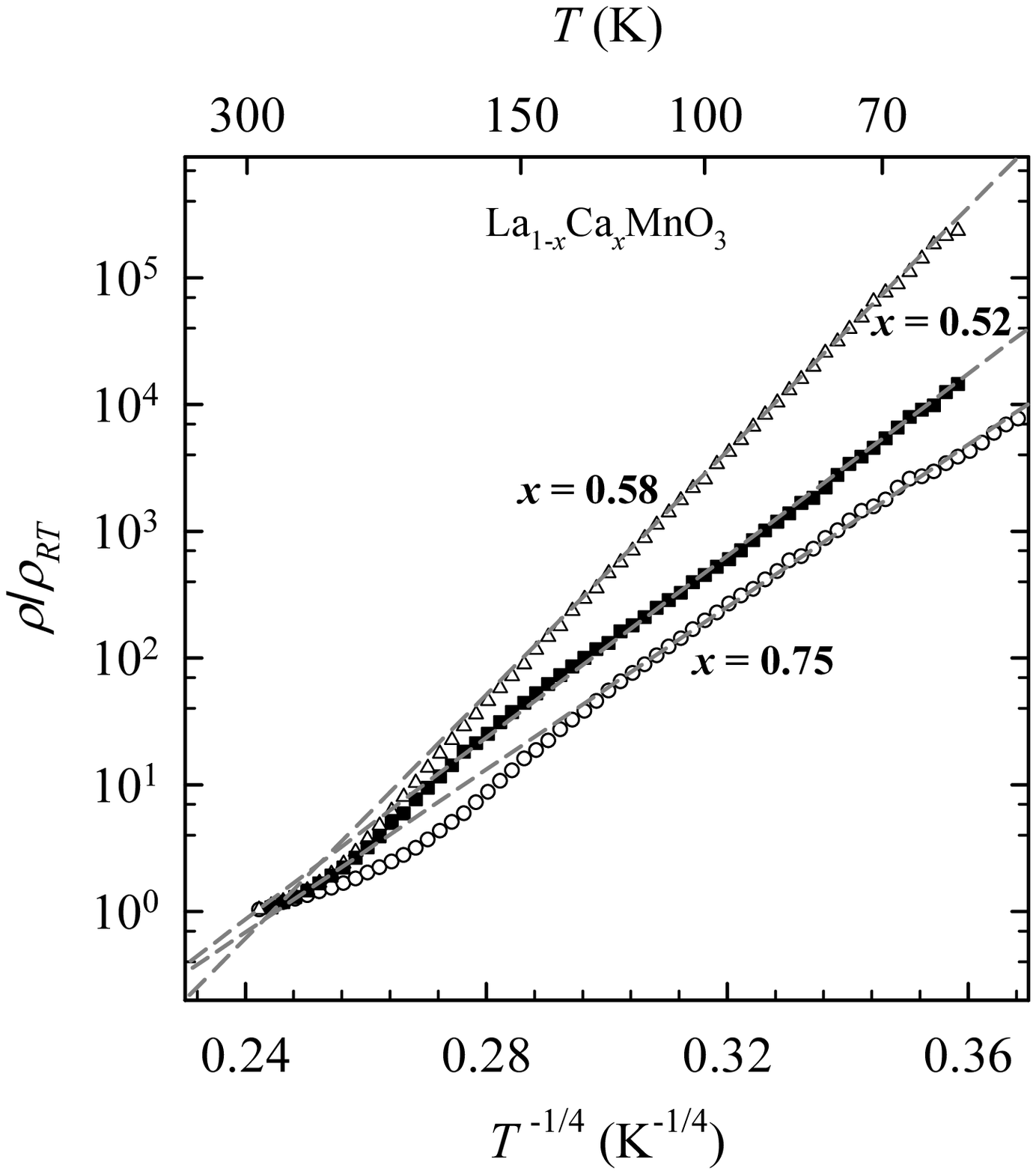}
\caption{$\rho/\rho_{RT}$ versus $T^{–1/4}$ for La$_{1–x}$Ca$_{x}$MnO$_{3}$, $x = 0.52$, $x = 0.58$ and $x = 0.75$. The dashed lines represent fits to the Mott's 3D VRH mechanism.}
\label{fig:VRH}
\end{figure}

In order to investigate the conduction mechanism below $T_{\mathrm{CO}}$ we have applied the Mott's variable-range hopping (VRH) model. This model has been used so far to describe the mechanism of conductivity for $x < 0.5$ in the temperature range above the metal--insulator transition \cite{viret}, for polycrystalline samples of La$_{1-x}$Ca$_{x}$MnO$_{3}$ ($0.5 \leq x < 1$) \cite{zhou}, for some other manganites like bi–layer manganite LaSr$_{2}$Mn$_{2}$O$_{7}$ \cite{matsukawa} or to describe the conduction mechanism in the CO state for the charge–ordered polycrystalline Pr$_{1–x}$Ca$_{x}$MnO$_{3}$ ($x > 0.5$) \cite{zheng}. According to the VRH model, the temperature dependence of resistivity is represented by the formula $\rho(T) = \rho_{\infty}\mathrm{exp}(T_{0}/T)^{1/(1+d)}$, where $\rho_{\infty}$  is a constant, \textit{d} denotes the dimensionality of the system and $T_{0}$ is the Mott's activation energy which is indicative of disorder, and thus of localization. $T_{0}$ is proportional to $1/[N(E_{F})\xi^{d}]$, where $N(E_{F})$ is the density of states at the Fermi level and $\xi$ is the localization length. Figure \ref{fig:VRH} shows the best fits for our samples with $x = 0.52$, 0.58 and 0.75 that were obtained for $d = 3$. In other words, the transport mechanism for $T < T_{\mathrm{CO}}$ can be attributed to the 3D Mott’s variable--range hopping conductivity. The fitting parameters are listed in Table \ref{tab} showing a similar order of magnitude of $T_{0}$ when compared with some previous reports \cite{viret,matsukawa,zheng}. Note that according to the Mott’s VRH theory the parameter $T_{0}$ is inversely proportional to the localization length of the carriers and the larger value of $T_{0}$ may imply the existence of a greater intrinsic disorder in a sample.

\begin{figure}
\centering
\includegraphics[width=0.7\textwidth]{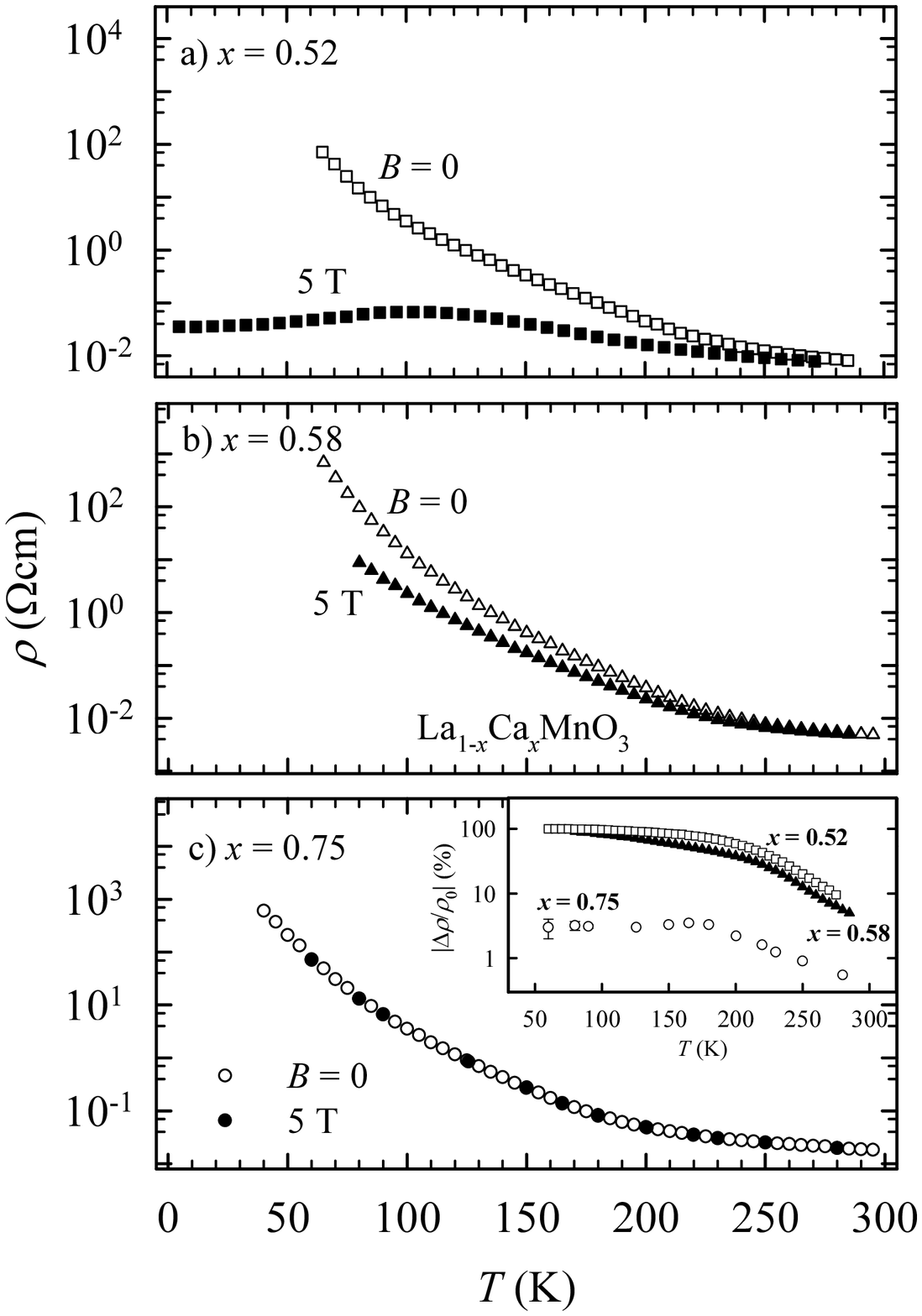}
\caption{Temperature dependence of resistivity for La$_{1–x}$Ca$_{x}$MnO$_{3}$ for $B = 0$ (empty symbols) and $B = 5$ T (full symbols): a) $x = 0.52$, b) $x = 0.58$, c) $x = 0.75$. The inset shows the temperature dependence of transversal magnetoresistance $\Delta\rho/\rho_{0}$ for 5 T: $x = 0.52$, $x = 0.58$ and $x = 0.75$.}
\label{fig:resistivity_5T}
\end{figure}

The applied external magnetic field significantly reduces the resistivity (negative magnetoresistance) and yields different temperature dependences. This is shown in Fig. \ref{fig:resistivity_5T}, where the temperature variations of the resistivity in the magnetic field $B = 0$ and $B = 5$ T for $x = 0.52$, 0.58 and 0.75  are plotted (the magnetic field is perpendicular to the \textit{ac} plane of the film and consequently to the current direction). It can be seen that the biggest reduction of the resistivity (for $B = 5$ T) is for $x = 0.52$ (Fig. \ref{fig:resistivity_5T}a) whereas for $x = 0.75$ the change of the resistivity is almost negligible (Fig. \ref{fig:resistivity_5T}c). Note that for $x = 0.52$ below $\approx 110$ K an insulator–metal transition appears and this will be discussed later. The influence of \textit{B} on $T_{\mathrm{CO}}$ can also be identified: $T_{\mathrm{CO}}$ decreases with increasing magnetic field; the change $\Delta T_{\mathrm{CO}} = T_{\mathrm{CO}}(0 ~\mathrm{T}) – T_{\mathrm{CO}}(5 ~\mathrm{T})$ is $\approx 9$ K for $x = 0.52$, and $\approx 5$~K for $x = 0.58$, whereas there is no change for $x = 0.75$. Our result is in agreement with Ref. \cite{zhou} where the shift of $T_{\mathrm{CO}}$ to lower temperatures is interpreted as the suppression of the CO state in a magnetic field and it was found that with increasing Mn$^{4+}$ content the value of $\Delta T_{\mathrm{CO}}$ decrease becomes negligible at $x = 0.75$, where the CO state is most stable. To see more closely the diversity of the magnetic field influence on different sample concentrations, we have plotted in the inset of Fig. \ref{fig:resistivity_5T}c the temperature variations of the absolute value of the magnetoresistance $\left|\Delta\rho/\rho_{0}\right|(\%)$ (defined as $\left|100 \times [\rho(B) – \rho(0)]/\rho(0)\right|$). While the magnetoresistance at 80 K reaches the values as high as  –99.5 \% for $x = 0.52$ and –91 \% for $x = 0.58$, it does not exceed –3 \% for $x = 0.75$. Equally important is that the temperature dependence of MR shows different slopes above and below $T_{\mathrm{CO}}$ for all three \textit{x}, and the temperature independent MR for $x = 0.75$ below $T_{\mathrm{CO}}$ demonstrates the stability of the CO state (which is not influenced by magnetic field up to 5 T for this concentration) \cite{li}. It is worth noting that some authors use another definition for magnetoresistance of these systems showing CMR which reads $\Delta\rho/\rho_{0} = 100 \times [\rho(0) – \rho(B)]/\rho(B)$ and which in our case gives the values at 80 K of 30,000~\%, 1000 \% and 4 \% for $x = 0.52$, 0.58 and 0.75, respectively. These high values show clearly that CMR effect is present for $x = 0.52$ and (slightly less) for $x = 0.58$, while disappears completely for $x = 0.75$. Regardless of the approach used for calculating the MR values, the same conclusion could be deduced that the ferromagnetic phase is present in the phase diagram for $x > 0.5$. In other words, we find the coexistence of charge–ordered insulating and ferromagnetic metallic phases for $0.5 < x \leq 0.58$. This behavior can be interpreted as the consequence of the inhomogeneities that arise from phase competition \cite{dagotto05,dagotto01,tokura,shenoy}. Namely, it is known that physical properties of manganites, despite their high chemical homogeneity, can show spatially inhomogeneous structures, i.e. regions showing different electronic orders. These differences actually mean spatially correlated arrangement of charge, spin, and/or orbitals that we can call clusters and are known as a phase separation. As it is widely accepted, the phase separation is very probably the key to the CMR phenomenon \cite{dagotto05,dagotto01,tokura,dagotto_science}, where the existence of preformed FM islands or clusters and their easy alignment with external magnetic fields cause a large magnetoresistance. In other words, some FM clusters persist along with the CO state for $x \geq 0.5$ (or FM regions over a CO/AF background). The volume fraction of these FM clusters becomes larger with the applied external magnetic field and for strong enough magnetic fields FM clusters can connect each other generating FM metallic regime in CO/AF state. This behavior was firstly found for $x = 0.5$ and termed as a ``melting'' of charge order by a magnetic field \cite{roy}. As mentioned previously, in an earlier study on the polycrystalline samples for $x = 0.5$, the resistivity showed metal–like behavior for $B > 12$ T \cite{zhou}. Figure \ref{fig:resistivity_5T}a shows evidence for this behavior for $x = 0.52$ and $B = 5$ T. This finding and the differences in strength of the magnetic field that induces metallic state also show that the conditions during sample preparations influence their physical properties as already shown in other systems \cite{dagotto05,tokura,belevtsev}.

\begin{figure}
\centering
\includegraphics[width=0.5\textwidth]{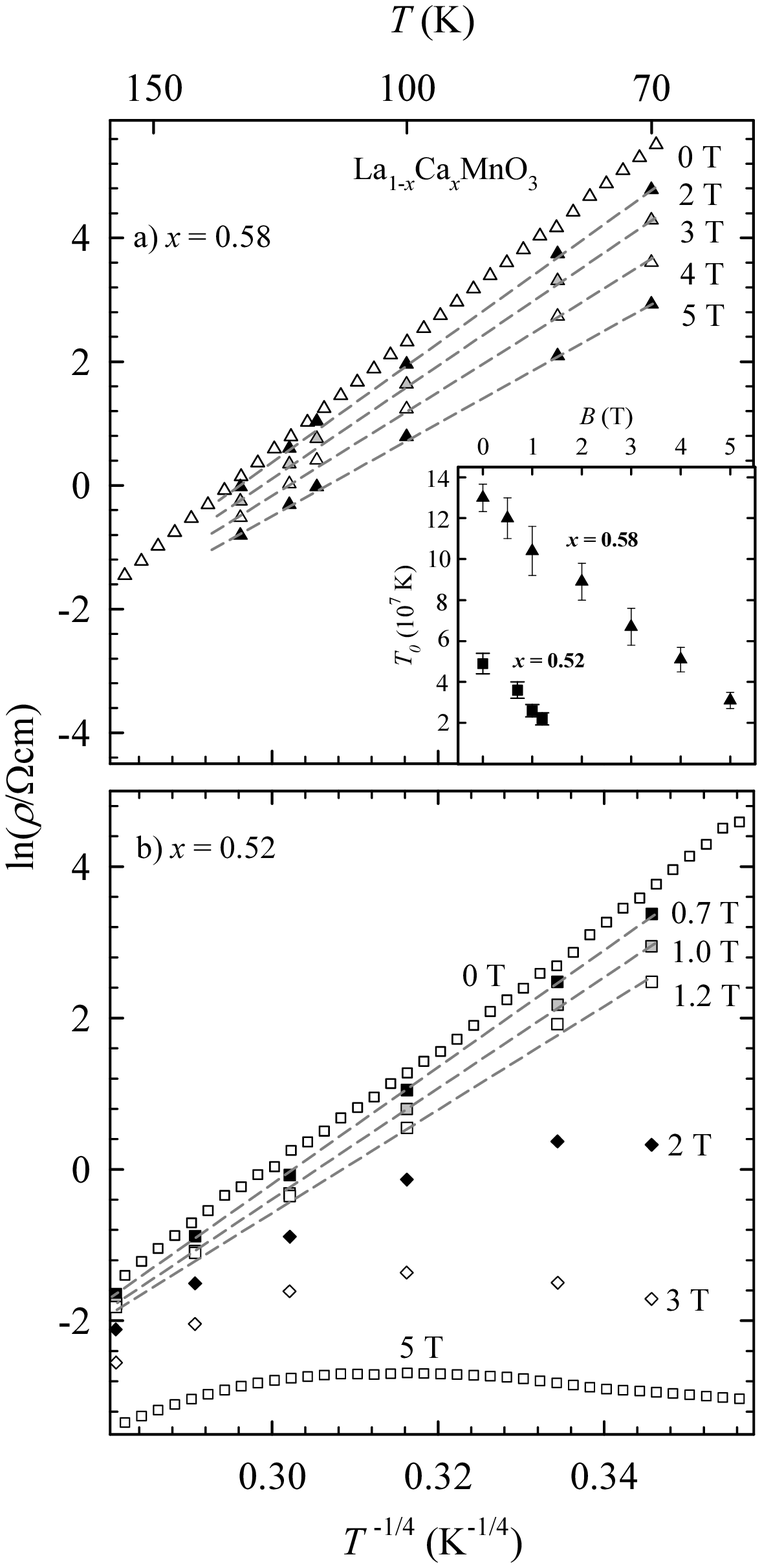}
\caption{ln$\rho$ versus $T^{–1/4}$ in different magnetic fields for La$_{1–x}$Ca$_{x}$MnO$_{3}$ a) $x = 0.58$, b) $x = 0.52$. The dashed lines represent fits to the Mott's 3D VRH mechanism. The inset shows the magnetic field dependence of the Mott's temperature $T_{0}$ for $x = 0.52$ (full squares) and $x = 0.58$ (full triangles).}
\label{fig:spin_VRH}
\end{figure}

According to our analysis the transport mechanism for $T < T_{\mathrm{CO}}$ in zero field is attributed to the Mott’s variable--range hopping conductivity for a 3D system, while the influence of external magnetic field on the hopping mechanism of conductivity can be monitored and investigated as shown in Fig. \ref{fig:spin_VRH}. The $T^{–1/4}$ dependences of resistivity for different values of the magnetic field ($0 \leq B \leq 5$ T) are presented for $x = 0.58$ (Fig. \ref{fig:spin_VRH}a) and $x = 0.52$ (Fig. \ref{fig:spin_VRH}b). For $x = 0.75$ the hopping conductivity shows a negligible change in magnetic field up to 5 T. On the other hand, for $0.5 < x \leq 0.58$ the application of a magnetic field alters the hopping parameters. The inset of Fig. \ref{fig:spin_VRH} shows magnetic field dependence of $T_{0}$ for $x = 0.52$ and 0.58, where the increase of \textit{B} causes a reduction of the characteristic temperature $T_{0}$. This $T_{0}$ decrease is due to a progressive spin alignment with the applied magnetic field, suggesting the delocalization of the electronic states. Applied magnetic field aligns the moments and reduces the disorder potential i.e. the degree of spin alignment gets increased by increasing \textit{B} \cite{dagotto01,tokura}. Spin dependent hopping as a modification of the Mott's VRH model was proposed by taking into account the dependence of the hopping barrier on the misorientation between the spins of electrons at an initial and a final state in an elementary hopping process \cite{wagner}. Also, in this model carriers are mobile and delocalized within the individual FM clusters, while the macroscopic resistivity is determined by the transfer of the carriers between neighboring spin clusters. The model was very satisfactorily applied in the semiconducting state and in the quasimetallic regime to interpret the temperature and field dependence of the resistivity in another CMR material Nd$_{0.52}$Sr$_{0.48}$MnO$_{3}$. Applying the same approach would imply that also in our case there exist FM regions over a CO/AF background. Indeed, for $x = 0.52$ and already for $B > 3$ T we noticed the tendency toward insulator–metal transition (Fig. \ref{fig:spin_VRH}b), while for $x = 0.58$ we did not find similar behavior up to 5 T (Fig. \ref{fig:spin_VRH}a). The value of $T_{0}$ ($B = 0$) for $x = 0.58$ is higher than for $x = 0.52$, but for both concentrations one observes a similar tendency of decrease with the increasing magnetic field. From this point of view, we cannot rule out the possibility that for $B > 5$ T and/or at lower temperatures the metal–insulator transition appears also for $x = 0.58$. On the other hand, for $x = 0.75$ a very small magnetoresistance without any fingerprints of CMR behavior, as well as the absence of magnetic field induced change in $T_{0}$ up to 5 T, exclude the presence of FM clusters in the sample. The declining influence of FM clusters in the CO/AF part of the phase diagram with increasing \textit{x} contributes to a possible explanation that the phase coexistence is the origin of the CMR phenomenon. 

\begin{figure}
\centering
\includegraphics[width=0.6\textwidth]{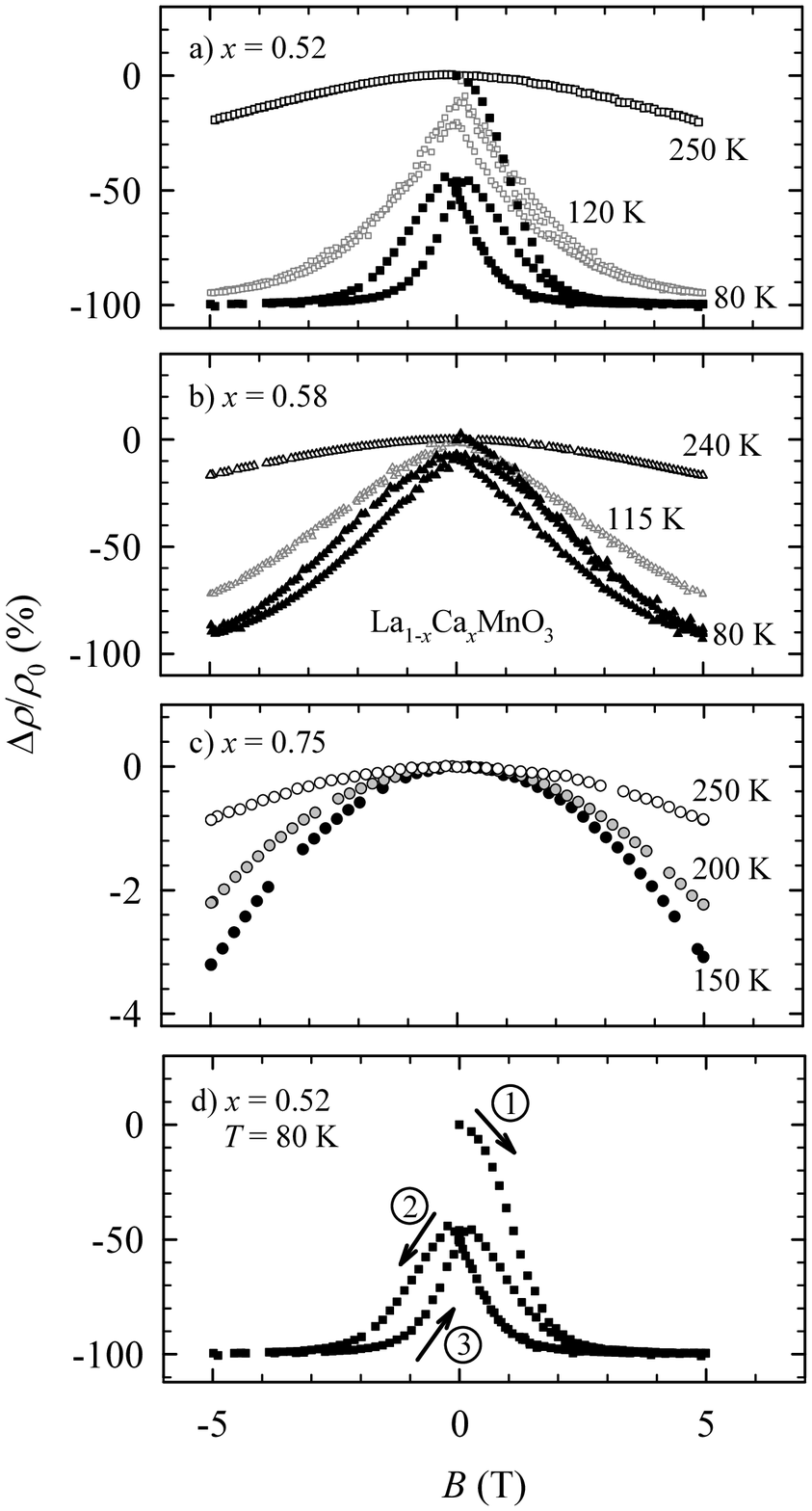}
\caption{Magnetic field dependence of transversal magnetoresistance $\Delta\rho/\rho_{0}$ at different temperatures for La$_{1–x}$Ca$_{x}$MnO$_{3}$, a) $x = 0.52$, b) $x = 0.58$, c) $x = 0.75$ and d) $x = 0.52$, $T = 80$ K (see text).}
\label{fig:MR}
\end{figure}

In order to further examine in detail the influence of magnetic field, the magnetoresistance for $0 \leq B \leq 5$ T was measured for all samples at several fixed temperatures. Fig. \ref{fig:MR} shows transversal magnetoresistance vs. magnetic field at several temperatures (above and below CO transition), for $x = 0.52$, 0.58 and 0.75. The difference in the $\Delta\rho/\rho_{0}$ field dependence is quite evident. Above the CO transition in the paramagnetic phase (at temperatures around 250 K for all samples) $\Delta\rho/\rho_{0}$ shows approximately $B^{2}$ behavior. For $x = 0.75$ the similar behavior is found at lower temperatures as well. On the other hand, for $x = 0.52$ and $0.58$ different $\Delta\rho/\rho_{0}$ vs. \textit{B} behavior is observed below $T_{\mathrm{CO}}$ (which is less pronounced for $x = 0.58$). A large drop in the resistivity due to the influence of the magnetic field is definitely a consequence of the magnetic field driven CO/AF towards FM transition indicating the presence of the CMR effect \cite{donnell}. Another phenomenon that we see at low temperatures, again more pronounced for $x = 0.52$ than $x = 0.58$ (and not present for $x = 0.75$), is the fact that the FM phase induced in the CO background state by an applied field, still persists after suppressing the field (Fig. \ref{fig:MR}d). As indicated by the arrows in Fig. \ref{fig:MR}d, after the initial sweep from 0 to 5 T (indicated as \raisebox{.5pt}{\textcircled{\raisebox{-.9pt} {1}}}), the magnetoresistance does not reach the same value when the field is swept back to $B = 0$. This point is indicated as \raisebox{.5pt}{\textcircled{\raisebox{-.9pt} {2}}} (the sweep from 5 T to $-5$ T) and \raisebox{.5pt}{\textcircled{\raisebox{-.9pt} {3}}} (the sweep from $-5$ T to 5 T). This could be due to the presence of non–equilibrium magnetic field induced (metamagnetic) clusters \cite{gosnet,shenoy}. This fact is an evidence of the presence of a phase coexistence meaning the presence of FM clusters over the CO/AF background for $0.52 \leq x \leq 0.58$. Moreover, the clearly seen magnetoresistance hysteresis around $B = 0$ presents the evidence of FM behavior. This point is qualitatively explained in the framework of the model that includes multiple domains with different magnitude and direction of the magnetization \cite{donnell}. The fact that all mentioned findings are very pronounced for $x = 0.52$ (Fig. \ref{fig:MR}a), present, but in a much less pronounced form, for $x = 0.58$ (Fig. \ref{fig:MR}b) and absent for $x = 0.75$ (Fig. \ref{fig:MR}c) could be attributed to the higher fraction of the FM clusters (or reduced fraction of the CO/AF phase) for $x = 0.52$ than $0.58$ and absence of the FM phase in $x = 0.75$. This conclusion is in concordance with our results presented so far.

The anisotropy of magnetoresistance (AMR) may be used to characterize the changes in the resistivity as the direction of the magnetic field \textit{B} changes with respect to the direction of the current \textit{j}, and our intention was to observe the possible presence of AMR behavior for $x > 0.5$. Several studies of the thin films of the CMR manganite La$_{1–x}$Ca$_{x}$MnO$_{3}$ with compositions within the range $x = 0.3 – 0.35$ revealed that both the in–plane ($j \parallel B$) and the out of plane ($j \perp B$) AMRs are present and are anomalous near the Curie temperature when compared to metallic ferromagnetic alloys \cite{donnell,li09,bibes,amaral}. The AMR in ferromagnetic alloys shows maximum when $j \parallel B$ \cite{campbell} and decreases monotonically with increasing temperature, while for CMR manganites it shows a peak at temperatures near the metal–insulator transition. Also, for the CMR manganites the AMR appears to be affected by the crystalline quality and the thickness of La$_{1–x}$Ca$_{x}$MnO$_{3}$ films deposited on substrates indicating an increase in the AMR amplitude with increasing an in–plane lattice strain with a decreasing thickness of the films \cite{gosnet,donnell98}. 

In conventional theories AMR in ferromagnetic alloys is associated with the magnetic state of the samples and due to the spin–orbit coupling it is a consequence of an anisotropic scattering of extended metallic conduction states \cite{malozemoff}. For CMR manganites a possible explanation is that AMR is due to a local spin–orbit induced, orbital deformation that influences the local hopping conduction mechanism which characterize the manganites \cite{donnell}. It was also shown that the domain rotation and magnetocrystalline anisotropy have an important influence on the AMR of manganites \cite{donnell,amaral}. However, all our samples have the same thickness and therefore we do not expect that the strain inhomogeneities (due to the substrate–film lattice interaction) should affect the AMR for different \textit{x} \cite{belevtsev01}. Our intention is to explore the possible change in the AMR with increasing \textit{x} for $0.52 \leq x \leq 0.75$ to monitor perceived phenomena of the CMR for $x > 0.5$. 

\begin{figure}
\centering
\includegraphics[width=0.6\textwidth]{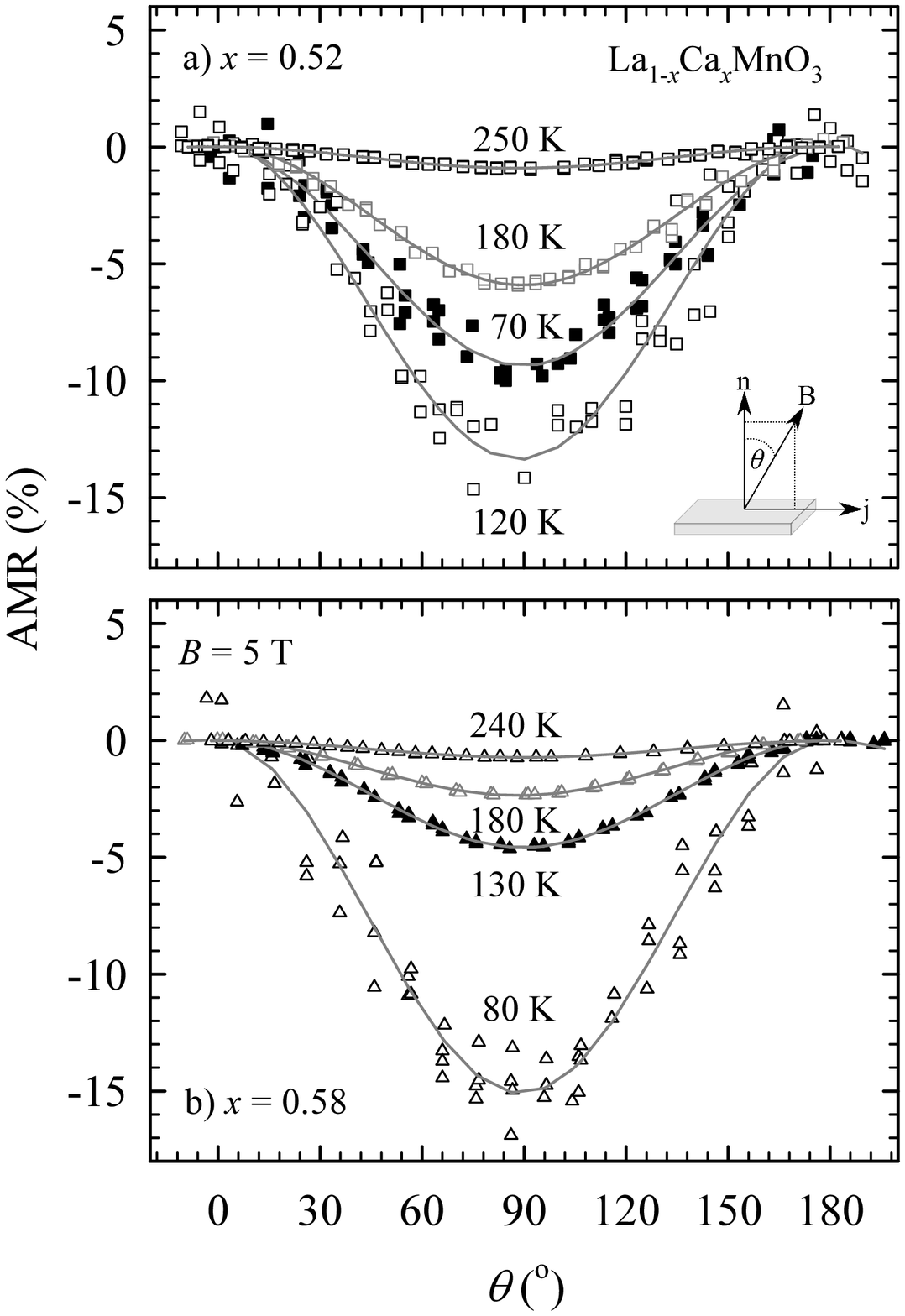}
\caption{Anglular dependence of magnetoresistance (AMR–see text) in 5 T at different temperatures for La$_{1–x}$Ca$_{x}$MnO$_{3}$, a) $x = 0.52$ and b) $x = 0.58$. Full lines show fits to $\mathrm{AMR} = \mathrm{AMR}_{\mathrm{max}}\mathrm{sin}^2(\theta)$. The sketch defines the angle $\theta$; $\theta = 0^{\circ}$ for transversal ($j \perp B$) and $\theta = 90^{\circ}$  for longitudinal 
magnetoresistance ($j \parallel B$).}
\label{fig:AMR}
\end{figure}
\begin{figure}
\centering
\includegraphics[width=0.6\textwidth]{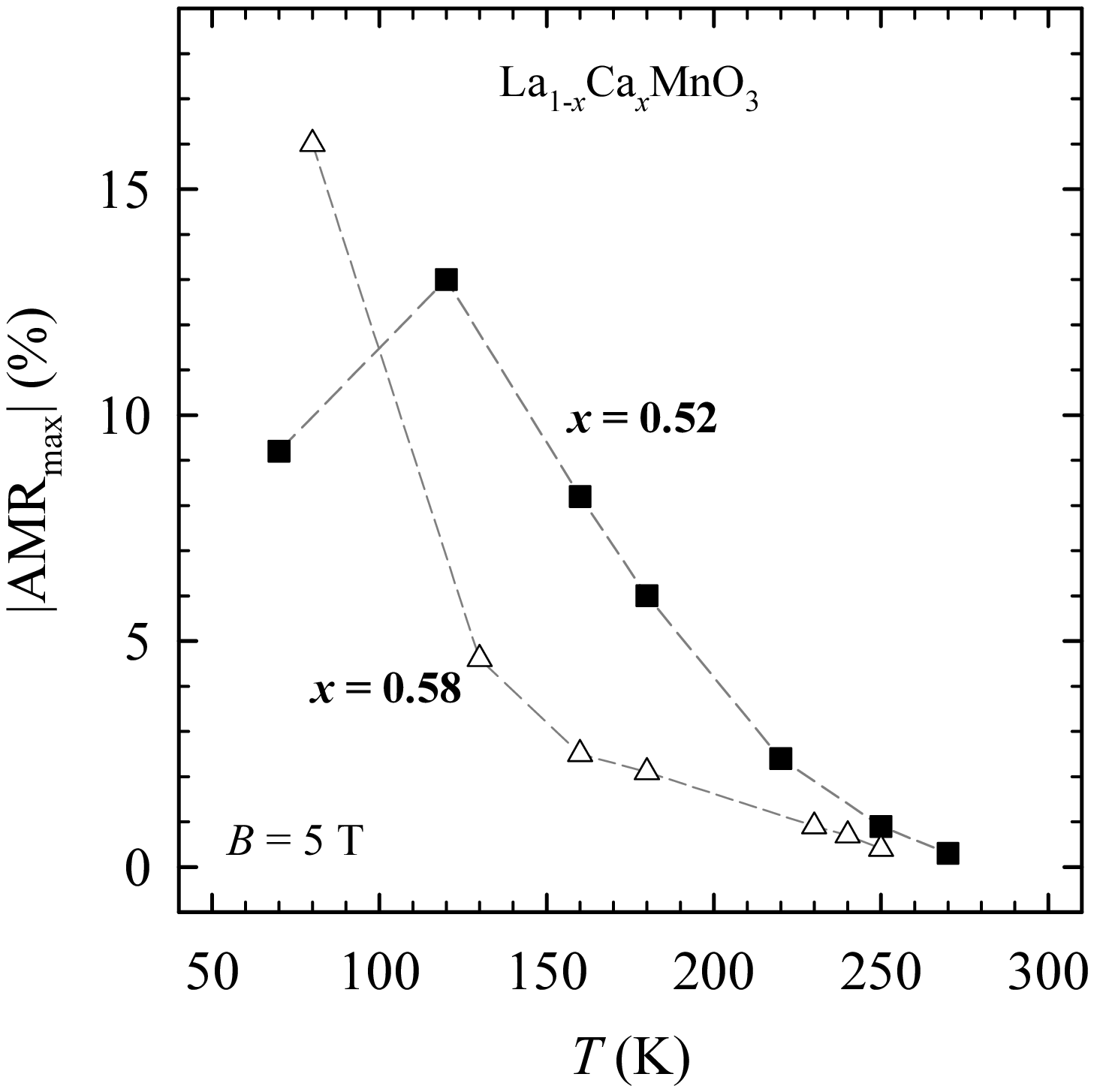}
\caption{Temperature dependence of $\left|\mathrm{AMR}_{\mathrm{max}}\right|$ in $B = 5$ T for La$_{1–x}$Ca$_{x}$MnO$_{3}$, $x = 0.52$ and $x = 0.58$. Dashed line is a guide for the eye.}
\label{fig:AMR_max}
\end{figure}

Figure \ref{fig:AMR} shows the angular dependence of resistivity in the magnetic field of 5 T defined as $\mathrm{AMR} = [\rho(5 ~\mathrm{T},\theta)–\rho(5 ~\mathrm{T},\theta=0^{\circ})]/\rho(5 ~\mathrm{T},\theta=0^{\circ})$ for $x = 0.52$ (Fig. \ref{fig:AMR}a) and $x = 0.58$ (Fig. \ref{fig:AMR}b). For $x = 0.75$ the AMR was not detected at all and therefore is not shown. As seen from the figures, the absolute value of the magnetoresistance for both $x = 0.52$ and $x = 0.58$ samples is larger (and reaches the maximum value) in the parallel configuration ($j \parallel B$) than in the perpendicular configuration ($j \perp B$). In the framework of the phenomenological model that includes a multidomain configuration of the sample, the AMR results from the changing populations of the three domain types (parallel, antiparallel and transverse) and their influence on the resistivity through AMR and CMR \cite{donnell,amaral,donnell98}. The lines in Figs. \ref{fig:AMR}a and \ref{fig:AMR}b denote fits to sin$^{2}\theta$ which is in agreement with the result for the CMR manganite with $x = 0.35$, obtained for the same direction of the field rotation \cite{egilmez}. The absolute values of maximum, AMR$_{\mathrm{max}}$, differ for $x = 0.52$ and $x = 0.58$, and Fig. \ref{fig:AMR_max} presents their temperature dependences for $B = 5$ T. For $T > 130$ K the $x = 0.52$ sample has higher values of AMR$_{\mathrm{max}}$ with respect to the values for $x = 0.58$; this may be attributed to a higher fraction of FM clusters in the sample, which is again in a good agreement with our previously shown results. In line with this, the magnetic field driven CO/AF insulator to FM metallic transition for $x = 0.52$ occurs below about 110 K (i.e. below the maximum of the AMR$_{\mathrm{max}}(T)$ curve for this sample). This point is also clearly supported by the large drop in the resistivity in the $\rho$(5 T) vs. \textit{T} for $x = 0.52$ presented in Fig. \ref{fig:resistivity_5T}a. On the other hand, for $x = 0.58$ the possibility for the insulator to FM metallic transition (at lower \textit{T} and higher \textit{B}) cannot be excluded, whereas for $x = 0.75$ all the data presented indicate that inside its CO/AF state there are no FM clusters that would affect its magnetotransport properties. 

\section{Conclusions}
We have studied magnetotransport properties of La$_{1-x}$Ca$_{x}$MnO$_{3}$ for the high Ca concentrations ($0.52 \leq x \leq 0.75$) in order to shed more light on the La$_{1–x}$Ca$_{x}$MnO$_{3}$ properties in the overdoped part of the phase diagram. The resistivity shows the semiconductor/insulator--like behavior in zero field for all three concentrations studied, obeying the 3D Mott's variable--range hopping model, at lower temperatures for $T < T_{\mathrm{CO}}$. The applied magnetic field strongly alters the hopping parameters for $x = 0.52$ and 0.58, implying that the spin–dependent hopping is significantly enhanced causing the delocalization of the electronic states. For $x = 0 .75$ the hopping parameters are almost not influenced by a magnetic field up to 5 T. Consequently, the magnetoresistance is negative; its absolute value is huge for $x = 0.52$ and somewhat reduced for $x = 0.58$, whereas it is almost negligible for $x = 0.75$. The magnetic field sweeps and the angular dependence of the magnetoresistance at fixed temperatures gave evidences of the presence of the CMR behavior for $x = 0.52$ and $x = 0.58$ and its absence for $x = 0.75$. Our resistivity and magnetoresistance data have been sucessfully analyzed within the model/picture that for $x \leq 0.58$ the CMR behavior is due to the strong competition of the existing FM metallic state regions and CO/AF background state, which additionally supports the approach that a phase separation is at the origin of the CMR phenomenon. Further investigations for different $x$, in higher magnetic fields and in extended temperature range could give additional information about the phase separation in this system.  

\section*{Acknowledgements}

This work was supported by the Croatian Science Foundation project IP-2013-11-1011. We thank B. Gorshunov for the very helpful discussions.

\end{document}